\documentclass[lettersize,journal]{IEEEtran}

\usepackage{amsmath,amsfonts,amssymb,amsthm}
\usepackage{algorithmic}
\usepackage{array}
\usepackage{textcomp}
\usepackage{stfloats}
\usepackage{url}
\usepackage{verbatim}
\usepackage{graphicx}
\usepackage{color, soul}
\usepackage{bm}
\usepackage{hyperref}
\usepackage{cite}
\usepackage{makecell,multirow}
\usepackage{subcaption}

\def\BibTeX{{\rm B\kern-.05em{\sc i\kern-.025em b}\kern-.08em
    T\kern-.1667em\lower.7ex\hbox{E}\kern-.125emX}}
\usepackage{balance}
\begin{document}
\title{User-centric Immersive Communications in 6G: A Data-oriented Framework via Digital Twin}

\author{    Conghao~Zhou,~\IEEEmembership{Member,~IEEE,}
            Shisheng~Hu,~\IEEEmembership{Student~Member,~IEEE,}
            Jie~Gao,~\IEEEmembership{Senior~Member,~IEEE,}
            Xinyu~Huang,~\IEEEmembership{Student~Member,~IEEE,}         
            Weihua~Zhuang,~\IEEEmembership{Fellow,~IEEE,}
            and Xuemin~(Sherman)~Shen,~\IEEEmembership{Fellow,~IEEE}

            \thanks{Conghao Zhou is with the School of Telecommunications Engineering, Xidian University, China. He was with the Department of Electrical and Computer Engineering, University of Waterloo, Waterloo, ON N2L 3G1, Canada (e-mail:~c89zhou@uwaterloo.ca).}

            \thanks{Shisheng~Hu (Corresponding Author), Xinyu~Huang, Weihua~Zhuang, and Xuemin (Sherman) Shen are with the Department of Electrical and Computer Engineering, University of Waterloo, Waterloo, ON N2L 3G1, Canada (e-mail:\{s97hu, x357huan, wzhuang, sshen\}@uwaterloo.ca).}

            \thanks{Jie~Gao is with the School of Information Technology, Carleton University, Ottawa, ON K1S 5B6, Canada (email:~jie.gao6@carleton.ca).}
        }


\maketitle

\begin{abstract}

In this article, we present a novel user-centric service provision for immersive communications (IC) in 6G to deal with the uncertainty of individual user behaviors while satisfying unique requirements on the quality of multi-sensory experience. To this end, we propose a data-oriented framework for network resource management, featuring personalized data management that can support network modeling tailored to different user demands. Our framework leverages the digital twin (DT) technique as a key enabler. Particularly, a DT is established for each user, and the data attributes in the DT are customized based on the characteristics of the user. The DT functions, corresponding to various data operations, are customized in the development, evaluation, and update of network models to meet unique user demands. A trace-driven case study demonstrates the effectiveness of our framework in achieving user-centric IC and the significance of personalized data management in 6G.

\end{abstract}


\section{Introduction}

Immersive communications (IC), envisioned to seamlessly bridge the physical and virtual worlds, have been recognized as a key use case of sixth-generation (6G) communication networks by IMT-2030 due to their vast potentials~\cite{shen2023toward}. In education, entertainment, healthcare, and many other realms, IC can have an immense impact that profoundly changes our ways of living. The popularity of the Metaverse concept, as a manifestation, is propelling the development of a lifelike virtual world that provides users with a deep immersion in social interactions. Meanwhile, the commercialization of wearable displays that can blend virtual content with the physical world is unlocking immersive three-dimensional (3D) video and audio experiences for more and more users. Considerable progress has also been made in transmitting haptic information over a communication link or network. Haptic devices such as gloves can enrich the communication experiences of users by supporting remote operations with haptic feedback. As a result, IC will enable people in different corners of the world to interact with each other and enjoy vivid visual, auditory, and haptic experiences as if they are inches apart.

To support multi-sensory immersive interactions among users worldwide, transcending networks beyond 5G in all performance metrics is essential~\cite{peng2024semantic}. Moreover, due to the human user-centric characteristics of IC applications, 6G networks need to introduce new dimensions of capability to meet the unprecedented demands of individual users, including the following three aspects. First, in IC applications, even a slight body movement of a user,~e.g., a gaze shift, may result in substantial network resource demand to maintain the sense of immersion. Due to the inherent uncertainty of user behaviors, networks must accurately characterize the impact of individual user behaviors on their network resource demands. Such characterization often relies on big data, thereby lacking scalability when managing detailed data from a massive number of users. Second, the prevailing quality of service (QoS) and quality of experience (QoE) metrics, based on a fixed set of data attributes, cannot accurately reflect user satisfaction in IC. Since multi-sensory experience is susceptible to dynamic changes in the physical world, a fixed set of data attributes as the QoE metrics may not precisely reflect user satisfaction. Therefore, an adaptable set of QoE metrics, based on the dynamic adjustment of data attributes, is essential for accurately reflecting user satisfaction in IC applications. Third, it is a consensus that artificial intelligence (AI) will be foundational for IC, both in network management and application development. Due to significant differences in user behavior patterns, concerns arise about whether AI models, e.g., deep neural networks (DNNs), trained on aggregated data from many users can perform optimally for each individual user~\cite{tao2024wireless}. Thus, networks need to support the customization of AI models for IC users. Due to the aforementioned reasons, understanding user behavior patterns, meeting unique user demands, and thereby enabling \emph{user-centric service provision}, are expected in 6G networks to support IC. 

In this article, we advocate a new \emph{data-oriented} framework for network resource management to achieve user-centric immersive communications (UCIC) in 6G. Different from conventional network management approaches which emphasize the process of network modeling, the data-oriented framework emphasizes the process of enhancing both the quality and the quantity of network data\footnote{By network data, we refer to data related to communication networks, encompassing infrastructure, network environments, and users.} through systematic and personalized data management. Based on this framework, network modeling can be customized for individual users and enable user-centric service provision. We leverage the digital twin (DT) technique as a key enabler of the data-oriented framework~\cite{yu2019cybertwin}. By creating digital replicas for individual IC user devices, we can customize user profiles to offer comprehensive user-related data for personalized network modeling, including the identification of personalized QoE metrics and characterization of user-specific behaviors. Additionally, we define various DT functions for data life-cycle management to adapt personalized network modeling to the dynamic changes in both the physical world and the user behaviors. A case study demonstrates the effectiveness of our data-oriented framework via DT in achieving UCIC.

\section{User-centric Immersive Communications}

In this section, we provide a brief introduction to immersive communications and share our vision of UCIC.

\subsection{Overview of Immersive Communications}

IC refer to communication and networking technologies that deliver lifelike experiences to users. Emerging IC applications include extended reality (XR), holographic communication, and haptic communication. We summarize features of IC applications as follows.

    \begin{itemize}

        \item \textbf{Multi-sensory}: IC applications involve multi-sensory perception, including visual, auditory, and haptic, each corresponding to distinct sensations, e.g., temperature and pressure. User devices serve as interfaces to transmit and process the multi-sensory information involved in IC.

        \item \textbf{Multi-module}: IC requires a synergy of technologies, e.g., device pose tracking and annotation rendering in the case of augmented reality. A dedicated module for each technology provides the corresponding functionality in enabling immersive interactions. These modules require network support in terms of communication, computing, sensing, and so on.

        \item \textbf{Human-in-the-loop}: IC applications are inherently human-in-the-loop since they depend on continuous, real-time user interactions, where individual user behaviors directly influence other users' responses. This makes the user experience heavily dependent on human factors.

        \item \textbf{AI-native}: IC in 6G will be tied to AI. Many modules, such as device pose prediction, are AI-driven. Furthermore, 6G networks are envisioned to be AI-native, featuring intelligent resource management for delivering immersive experiences. For example, future 6G networks can leverage generative AI to create virtual content on user devices to reduce the communication load in content delivery.

    \end{itemize}

\subsection{Features of UCIC and Challenges}

UCIC is expected to provide users with immersive experiences while accommodating the diverse characteristics of users in IC applications. The core concept is to enable personalized service provisioning for each individual user via network resource management, including key aspects as follows. 

\begin{itemize}

        \item \textbf{Adaptable user QoE}: Since user satisfaction in IC applications is influenced by dynamic changes in factors such as user behaviors, UCIC should support continuous and automated adjustment of user QoE metrics based on the online analysis of user-related data. However, related research independent of domain-specific knowledge~\emph{a priori} remains in its early stages.

        \item \textbf{User-customized AI}: Given the differences in user behavior and user-perceived experience, UCIC should ensure that the AI-based models perform optimally for individual users, rather than over a large set of users on average. However, effectively leveraging data resources within networks to guide the training, inference, and updating of AI-based models for each individual user remains a challenge.  

        \item \textbf{Scalable user characterization}: Characterizing user demands, e.g., quantifying the impact of user behaviors on resource demands, is indispensable for on-demand network resource management in UCIC. In practice, differences among users may not lead to substantial variations in resource demands. Moreover, conducting big data analysis for each individual user can yield significant data management overhead~\cite{sun2024knowledge}. Identifying a proper set of users to characterize their unique demands in various network scenarios is essential yet challenging.

    \end{itemize}

\section{Data-oriented Network Resource Management}

In this section, we introduce the concept and the scope of data-oriented network resource management for UCIC in 6G.

\subsection{Concept of the Data-oriented Framework}

    \begin{figure}[t]
        \centering
        \includegraphics[width=0.49\textwidth]{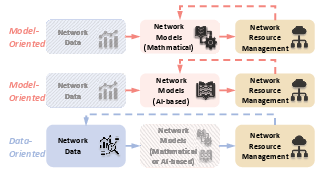}
        \caption{Model-oriented (including mathematical and AI-based) network resource management and a data-oriented framework.}\label{fig-data-centric}
    \end{figure}

The data-oriented concept originates from a field of AI, i.e., data-centric AI, which emphasizes data engineering for improving the performance of AI models~\cite{mazumder2023dataperf}. Extending this concept from AI to the communication network field, we advocate the new data-oriented framework for network resource management to support UCIC.

Researchers in the communication network field usually address network resource management problems by developing advanced \emph{network models}, initially mathematical models derived from observations. However, as networks become increasingly complex, large-scale, and heterogeneous, and services demand more diversified, developing accurate closed-form mathematical models becomes infeasible~\cite{shen2021holistic}. With the advent of AI, researchers can now extract information from massive network data using AI techniques, enabling network resource management decision making without relying on closed-form mathematical network models. In this context, AI can play the role of a network model in network resource management. 

As represented by the red blocks and arrows in Fig.~\ref{fig-data-centric}, \emph{model-oriented} network resource management focuses on developing mathematical or AI-based network models to extract information from network data for optimizing network resource management. Model-oriented network resource management may struggle to address issues such as missing and biased data, and cannot provide an adequate set of network data in terms of data volume, granularity, diversity, and other factors. Therefore, we propose a \emph{data-oriented} framework that emphasizes systematic data management to \emph{enhance the quality and quantity of network data}, thereby facilitating network modeling through mathematical, AI-based, or hybrid methods~\cite{shen2021holistic}. The proposed framework is represented by the blue blocks and arrows in Fig.~\ref{fig-data-centric}. 

\subsection{Scope of the Data-oriented Framework}\label{sec22}

The data-oriented framework can support UCIC in two aspects: 1) \emph{Enhancing} network data, which involves systematically improving the data quality across all phases of network modeling; 2) \emph{Expanding} network data, which involves introducing data life-cycle management to identify and utilize untapped data resources in existing network management. We elaborate on these two aspects via data operation tasks in three phases of network modeling, i.e., \emph{development}, \emph{evaluation}, and \emph{update} phases. Each data operation task deals with a specific data-related issue. In Table~\ref{table1}, we list typical data operation tasks in the scope of data-oriented network resource management. 

    \begin{table*}[ht]
        \footnotesize 
        \centering
        \captionsetup{justification=centering,singlelinecheck=false}
        \caption{Data operation tasks in data-oriented network resource management.}\label{table1}
            \begin{tabular}{c|c|c}
            \hline
            \hline
            \textbf{Phase}                                                  & \textbf{Data Operation Task} & \textbf{Description} \\ 
            \hline
            \hline
            \multirow{4}{*}{\shortstack{\textit{Network Model} \\ \textit{Development}}} & Data Preparation                 &  Identify potential data attributes and define the raw data collection process.            \\ \cline{2-3} 
                                                                           & Data Cleaning                 &  Transform raw data into a format appropriate for modeling.                    \\ \cline{2-3} 
                                                                           & Data Reduction                 &  Reduce feature dimension or sample size to lower modeling complexity.                     \\ \cline{2-3} 
                                                                           & Data Augmentation                 &  Generate variations of the existing data without additional collection.                 \\ \hline

            \multirow{2}{*}{\shortstack{\textit{Network Model} \\ \textit{Evaluation}}}           & In-distribution Data Generation        & Construct datasets for fine-grained in-distribution evaluation.                      \\ \cline{2-3} 
                                                                           & Out-of-distribution Data Generation                 &Generate out-of-distribution data to evaluate models in unexpected scenarios.                  \\ \hline 
                                                                           
            \multirow{3}{*}{\shortstack{\textit{Network Model} \\ \textit{Update}}}               & Data Valuation                 & Value the contribution of data to the modeling performance.                    \\ \cline{2-3} 
                                                                           & Data Quality Assurance               & Continuously measure, monitor, and ensure data quality in network management.                     \\ \cline{2-3} 
                                                                           & Resource Optimization             &  Adjust network resources allocated for data operation tasks.                    \\ 
                                                                           \hline
                                                                           \hline
            \end{tabular}
    \end{table*}

\subsubsection{Network Model Development Phase}

Developing network models, e.g., a QoE model, is fundamental for optimizing network resource management. In this phase, data operation tasks can be designed to prepare high-quality data to facilitate the development of network models with comprehensive information. For example, in IC, selecting an appropriate set of data attributes for identifying user QoE and defining the proper granularity of data samples can improve the accuracy of QoE-based resource demand prediction. Such data operation tasks can also be based on mathematical network models, which can provide a priori knowledge for efficient data management.

\subsubsection{Network Model Evaluation Phase} 

Succeeding network model development, the evaluation phase quantitatively characterizes the performance of network resource management decisions based on the developed network models. In this phase, data operation tasks focus on enhancing data quantity by generating additional data samples with diverse features. For example, to evaluate whether a data traffic model developed from aggregated data of multiple user devices meets the service demand of each individual user, data operation tasks in this phase should generate user-specific datasets to evaluate this user-agnostic model. These tasks support fine-grained evaluation of both closed-form mathematical network models and implicit AI-based network models from a data perspective, thereby guiding decision making for user-centric service provision. 

\subsubsection{Network Model Update Phase}

Due to the continuous changes of network environments, user behaviors, and other factors, regularly refining the network models for network resource management is necessary. Accordingly, the data required for updating those models should be continuously collected. The data operation tasks in the update phase concentrate on processing data for information extraction and timely triggering network model updates from a data perspective~\cite{zha2023data}. For example, by establishing quantitative measurements to characterize the impact of the data samples used for decision making on network performance, new factors, e.g., data distribution drift, can be introduced to trigger updates of AI-based network models used for network optimization.

\section{Digital Twin as An Enabler}

In this section, we introduce our data-oriented framework, leveraging the DT technique as its enabler. DTs serve as user profiles for individual user devices, facilitating the analysis of user data to enable UCIC.   

\subsection{Advantages of Digital Twins}

   For data-oriented network resource management, we adopt the DT technique as a key enabler for the following benefits.
    \begin{itemize}
        \item \textbf{Increase of available data attributes}: From the data quantity perspective, the DT technique can enrich the data attributes available for optimizing network resource management by establishing digital replicas of network entities. For example, to comprehensively depict the user characteristics, DTs established for individual user devices should incorporate application-specific data attributes, such as users' multi-sensory QoE requirements and pose variation patterns, which have not been adequately leveraged in existing network resource management~\cite{zhou2024digital}.

        \item \textbf{Customizable data updates}: From the data quality perspective, the DT technique supports flexible updating mechanisms for different data attributes, since the real-time synchronization between DTs and their corresponding physical objects ensures the availability of high-quality data for updates, allowing for dynamic adjustments of data quality as needed. As a result, the DT technique can facilitate the network model adaptation by adjusting network data quality, such as feature dimension and granularity. 

        \item \textbf{End-to-end evaluation capabilities}: From the data management perspective, the DT technique enables end-to-end evaluation of how data operations impact network performance or use satisfaction through network modeling and network resource management decision making. This is because DTs can receive feedback from physical objects, such as reports from user devices on delay performance, which is influenced by both network resource management and data operations. As a result, the DT technique introduces the possibility of guiding the joint management of network data and network resources.  

        \item \textbf{Data-model integration}: The DT technique facilitates the integration of network data and models. Mathematical network models can be used to generate network data via emulating various network scenarios, while network data can be used to derive network models via extracting patterns. The DT technique can incorporate mathematical and AI-based methods in network resource management.

    \end{itemize}

\subsection{Conceptual Architecture}

    \begin{figure*}[t]
        \centering
        \includegraphics[width=0.95\textwidth]{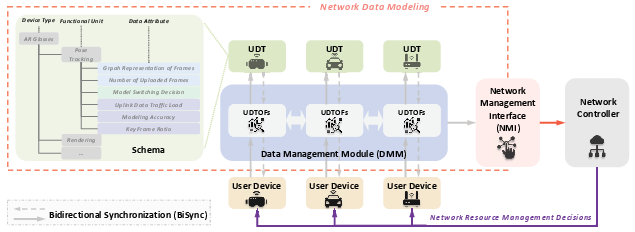}
        \caption{The conceptual architecture of data-oriented network resource management framework via UDTs.}\label{udt-arch}
    \end{figure*}

Toward user-centric service provision, our data-oriented framework concentrates on establishing user DTs for individual user devices and integrating user-related data management functionalities into network resource management. Building on our previous research~\cite{shen2021holistic}, we present the conceptual architecture of our framework in Fig.~\ref{udt-arch}. Rather than simply collecting data, the essence of our framework is developing a \emph{data model}, which organizes data and effectively supports the extraction of information~\cite{ramakrishnan2002database,ma2024one}. We refer to the process of customizing a data model to organize and process network data used for network resource management as \emph{network data modeling}, illustrated in the dashed red block in Fig.~\ref{udt-arch}. The four primary components of the architecture are outlined into follows:

    \begin{itemize}
        \item \textbf{User Digital Twin (UDT)}: We propose establishing a digital replica, termed UDT, for each individual user device under consideration. Within a UDT, we define a customizable \emph{schema} that describes the relationships among a set of data attributes associated with the corresponding user device for a target application~\cite{ramakrishnan2002database}, such as device ID, device type, and user's QoE requirements. The data attributes are intended to efficiently structure data samples for providing network resource management with explicit or implicit information. In addition, the schema specifies a data structure for efficient user data organization. Fig.~\ref{udt-arch} shows a reference schema customized for an augmented reality device to support device pose tracking. The schema organizes data attributes using a hierarchical data structure and specifies different data update mechanisms for different data attributes~\cite{zhou2024digital}. To ensure UDT feasibility, the schema should constrain data samples from a network perspective. For example, the granularity of channel condition data should align with that recorded by the access point. Additionally, privacy issues, such as unconscionable behavioral profiling and improper uses of user profiles, should be addressed when managing data containing user preference information~\cite{shen2023toward}.

        \item \textbf{Data Management Module (DMM)}: The DMM is responsible for executing various data operation tasks as mentioned in Subsection~\ref{sec22}. Specifically, within the DMM, we propose a set of \emph{UDT operation functions} (UDTOFs), each corresponding to a specific data operation task, as exemplified in Table~\ref{table1}. Data exchange between different UDTOFs is designed to support network-level data engineering while reducing redundant data operations. We classify UDTOFs into three categories:~I) functions contributing to the definition of schemas;~II) functions enabling bidirectional synchronization between UDTs and user devices; and~III) functions facilitating the development, evaluation, and updating of network models.

        \item \textbf{Bidirectional Synchronization (BiSync)}: The BiSync implements the synchronization between each user device and its corresponding UDT. From a user device to its UDT, there is a data flow starting from raw data collection at user devices, followed by data processing at the DMM, as indicated by the solid gray arrows in Fig.~\ref{udt-arch}. In the other direction, the synchronization is not simply transmitting data from a UDT to its corresponding user. Instead, UDTs influence network modeling and the decision making of network resource management, which in turn affects user devices. We indicate the synchronization in this direction with dashed gray arrows in Fig.~\ref{udt-arch}.

        \item \textbf{Network Management Interface (NMI)}: The NMI bridges data management and network resource management, with two primary functions. First, by utilizing the information from UDTOFs in the DMM, the NMI is responsible for the development, evaluation, and updating of network models, such as closed-form Poisson models or inherent network models contained in well-trained DNNs. Second, based on the network models, the NMI offers comprehensive information, e.g., estimated user QoE, for making network resource management decisions. These two functions may be integrated, particularly in AI-based approaches to network resource management where DNNs directly output the decisions.

    \end{itemize}

Via the above designs, the schema for a UDT organizes the relationships among the associated data attributes, serving as a tool to capture the \emph{static} characteristics of the corresponding user device. Meanwhile, UDTOFs in the DMM define rules of analyzing time-varying data values, serving as tools to capture the \emph{dynamic} characteristics of the user device. Since each UDTOF can be mathematical, AI-based, or hybrid, the NMI can flexibly coordinate network data modeling with network resource management decision making, especially via emerging hybrid-model-data-driven methods.

\subsection{Adaptive Network Data Modeling} 

    \begin{figure*}[t]
        \centering
        \includegraphics[width=0.85\textwidth]{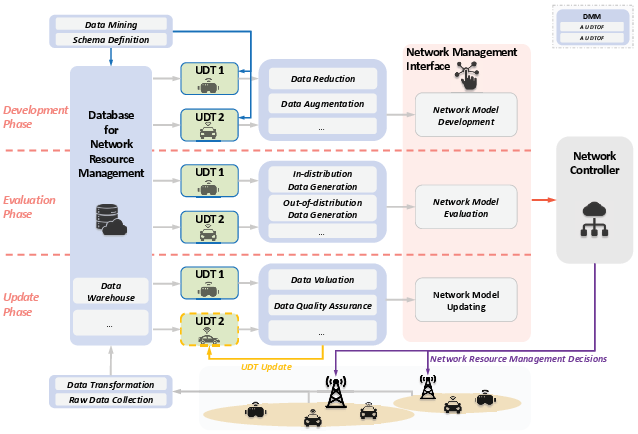}
        \caption{The illustration of data life-cycle management centered on UDTs.}\label{fig-life-cycle}
    \end{figure*}

Here, we discuss data life-cycle management for adapting the data model to network dynamics, thereby facilitating adaptive network resource management. We present the workflow corresponding to the development, evaluation, and update of the network model, which were briefly introduced in Subsection~\ref{sec22}.

\subsubsection{Development} 

In this phase, UDTOFs support network model development from the schema definition and data operation perspectives. To ensure the UDTs contain comprehensive data attributes while avoiding high costs from large database administration such as data warehouse, we consider a two-tier database as shown in Fig.~\ref{fig-life-cycle}. UDTOFs define the schema of a large database for low-redundancy data management across the entire network. Following this, they define the initial schemas of small databases for UDTs and enable UDTs to efficiently leverage the large database for ephemeral and customizable data management tailored to individual user devices. Since UDTOFs determine the overall set of data attributes available for QoE modeling, they are crucial for user QoE satisfaction. Thus, we can introduce data mining,~e.g.,~association rule learning or ablation study~\cite{wu2023characterizing}, into UDTOFs to precisely identify users' specific QoE metrics and corresponding data attributes for QoE modeling.

UDTOFs concentrate on preparing high-quantity data to enhance the capture of information, thereby facilitating the development of network models, particularly for AI-based QoE modeling. For example, they support QoE modeling by uncovering latent relationships between allocated radio resources and the resulting device pose tracking errors in XR~\cite{zhou2024digital}. Typical data operation tasks for UDTOFs in this phase are outlined in Table~\ref{table1}. Data reduction enhances efficiency and accuracy in network model development by removing irrelevant or redundant data, while data augmentation improves the generalization capabilities of network models by enhancing data diversity.

\subsubsection{Evaluation} 

UDTOFs in this phase mainly support the \emph{fine-grained evaluation} of network model performance from two perspectives. First, by integrating mathematical network models with network data, they can emulate atypical network scenarios not captured during the development phase, thereby evaluating the robustness of a network model. Second, UDTOFs can evaluate the performance of individual user devices under resource management decisions based on a general network model, e.g., a user-averaged data traffic model. Techniques such as generative AI or data slicing, which partition datasets involving multiple user devices into datasets specific to each user device~\cite{larose2014discovering}, can be applied for this purpose. From these two aspects, UDTOFs can exploit additional network data to increase the granularity of network model evaluation, thereby facilitating fine-grained network resource management.   
 
\subsubsection{Update} 

In this phase, UDTOFs facilitate updating both network models and UDTs. Regarding network model updates, a conventional approach is to trigger reactive updates by continuously monitoring and analyzing network performance data, which can be viewed as a downstream analysis of network modeling. Using UDTs, one can update network models through proactive assessment of the quality of the data input into the network models, representing an upstream analysis. Specifically, new metrics including the data quality of a set of data samples,~e.g., data distribution shifts, or the value of individual data samples,~e.g., the Shapely value, can be employed to proactively initiate network model updates~\cite{ghorbani2020distributional}. This can address certain issues related to network modeling at an early stage. 

Since UDTOFs are centered on UDTs, the evolution of UDTs, specifically UDT schema update, becomes necessary to capture the changing properties of user devices using real-time data. To this end, the tasks of data collection, transformation, and analysis should be properly defined for UDTOFs to ensure that the updated schema provides comprehensive information to update network models. Given the additional consumption of network resources, e.g., data storage and computing power, associated with UDTOFs, it is essential to strike a balance between the costs of UDTOFs and the resulting network performance gains. 

\section{Case Study}

In this case study, we show the effectiveness of our UDT-based framework in supporting user-centric virtual content delivery for the IC application of mobile augmented reality (MAR)~\cite{zhou2024digital,han2020vivo}.

\subsection{Considered Scenario}

In the considered scenario, MAR users within the coverage of an AP move freely with six degrees of freedom. An edge server deployed at the AP delivers volumetric videos as virtual content to MAR devices, allowing seamless integration with the real-world environment surrounding the MAR users. Each frame of a volumetric video is composed of a dense point cloud, posing a challenge in meeting the stringent latency requirement for MAR. A classic approach involves spatially segmenting the point cloud in each frame into multiple tiles and proactively delivering only the tiles that will likely fall within the field of view of an MAR user, according to the prediction of the MAR device pose. As the edge server needs to collect data related to device poses for pose prediction, the network should allocate sufficient uplink communication resources to ensure timely pose data collection. We use a public dataset that records the pose traces of $40$ users watching a volumetric video (\url{https://github.com/Yong-Chen94/6DoF_Video_FoV_Dataset}, titled \textit{Longdress}).

\subsection{Performance Evaluation}

To validate the effectiveness of the proposed data-oriented framework, this case study focuses on QoE-oriented service provision for MAR since QoE plays a critical role in assessing the level of user immersion in MAR. 

\begin{figure}
    \centering

    \subfloat[Comparison of proposed UDT-based and user-agnostic approaches to QoE modeling.]{\includegraphics[width=0.48\textwidth]{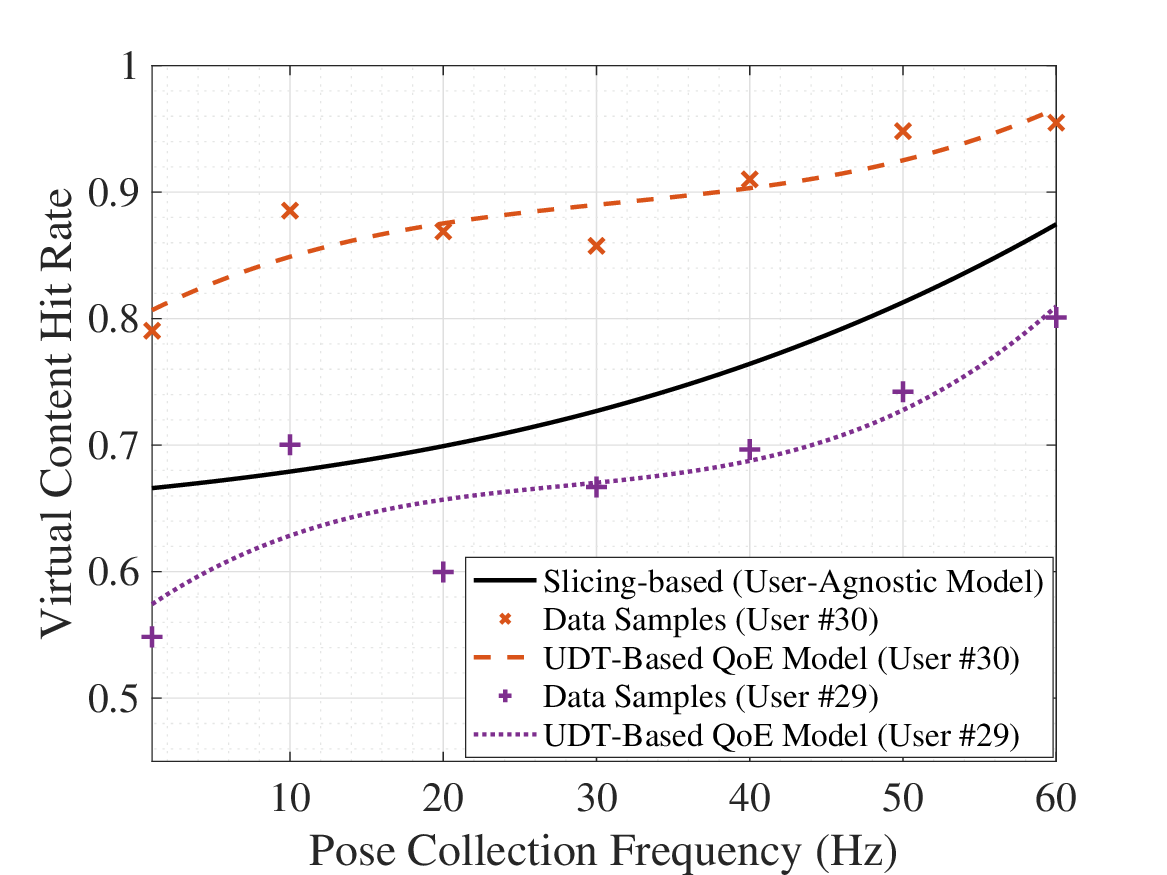}\label{simulation_fig2}}

    \subfloat[The impact of the number of established UDTs on the average QoE modeling error.]{\includegraphics[width=0.48\textwidth]{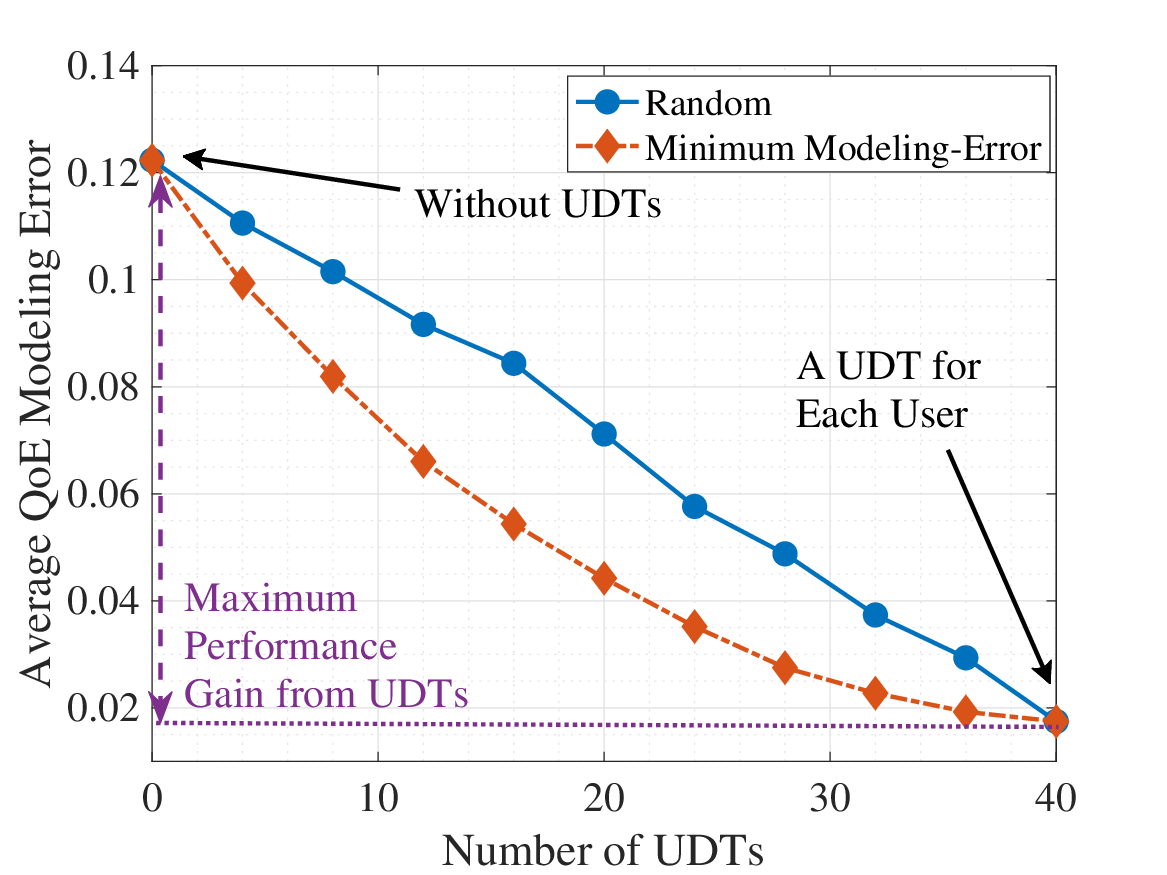}\label{simulation_fig3}}

    \caption{Effectiveness of UDTs in supporting virtual content delivery for MAR.}

    \label{eachmetric}
\end{figure}

\subsubsection{User QoE Modeling}

To achieve satisfactory QoE, we establish UDTs to help build QoE models that map from network resource management decisions to corresponding MAR users' QoE. We adopt a hierarchical data structure for the user profile in each UDT, as outlined in our prior work~\cite{zhou2024digital}. In addition to device category, user ID, and timestamps, the user profile incorporates two key data attributes for QoE modeling. First, the virtual content hit rate (VCHR) is used as a QoE metric, representing the ratio of overlapped point cloud between the delivered tiles and the rendered tiles, since it impacts user immersion in MAR~\cite{han2020vivo}. Second, the pose data collection frequency is employed to represent a network resource management decision, assuming an identical uplink resource consumption for each pose data collection. Following this data structure, data samples are prepared for modeling the relation between the pose collection frequency and the user QoE, i.e., the VCHR. The third-order polynomial regression is used to develop a QoE model based on the data samples in each UDT. 

In Fig.~\ref{simulation_fig2}, we show the VCHR with different pose collection frequencies. For comparison, we adopt a network slicing-based management approach commonly employed in 5G networks, as a baseline, which builds a single \textit{user-agnostic} QoE model based on the data samples collected from all MAR users. The VCHRs of all users increase with the pose collection frequency. Nevertheless, the trend varies drastically with users. For example, the VCHR of User~\#$29$ is lower and increases more drastically than User~\#$30$. This is because the users exhibit different initial poses and different patterns in changing their poses in watching the volumetric video. In Fig.~\ref{simulation_fig2}, we show the results of the user-agnostic model to characterize the relation. The user-agnostic model exhibits a significant error in modeling the QoE of individual users, especially for User~\#$30$. By comparison, establishing UDTs for user-specific QoE modeling enhances the QoE modeling accuracy. Leveraging user QoE information, our framework can be extended to QoE-oriented service provision for a broader range of 6G applications beyond IC.

\subsubsection{MAR User Selection for UDT Establishment}

While establishing a UDT benefits user-specific QoE modeling, the associated costs cannot be overlooked. Specifically, the DMM must consume network resources to support data operations for developing and using a QoE model. Therefore, it is important to select an appropriate set of MAR users for UDT establishment to improve QoE modeling accuracy while satisfying network resource constraints. Given the number of MAR users to be selected for UDT establishment, we compare two approaches to MAR user selection for UDT establishment, i.e., (i) \textit{random}: randomly selecting the MAR users; and (ii) \textit{minimum modeling-error}: calculating the modeling error for each user when the user-agnostic model is adopted for this MAR user and then selecting the users with the highest modeling errors. For users without UDTs, the user-agnostic QoE model is applied.

In Fig.~\ref{simulation_fig3}, given different numbers of UDTs established, we show the QoE modeling error averaged over all MAR users. When the number of UDTs is set to $0$ and $40$, UDTs are established for none and all of the MAR users, respectively. With the UDTs, the average modeling error largely decreases. To enhance scalability, both approaches can reduce signaling overhead and energy consumption by selecting user devices for UDT establishment. We observe that the minimum modeling-error approach achieves higher modeling accuracy. Based on this observation, properly balancing performance gains with the associated costs of UDT establishment is crucial for optimizing 6G networks.

\section{Conclusion}

In this article, we have introduced a data-oriented network resource management framework to facilitate user-centric immersive communications. The framework can support personalized data management for user-centric service provision by leveraging the DT technique. Our research lays the groundwork for integrating data management and resource management to trailblaze network automation in 6G. In the future, we will investigate closed-loop automated network resource management to enable scalable user-centric service provision for immersive communications.

\bibliography{ref}

\begin{thebibliography}{10}
\providecommand{\url}[1]{#1}
\csname url@samestyle\endcsname
\providecommand{\newblock}{\relax}
\providecommand{\bibinfo}[2]{#2}
\providecommand{\BIBentrySTDinterwordspacing}{\spaceskip=0pt\relax}
\providecommand{\BIBentryALTinterwordstretchfactor}{4}
\providecommand{\BIBentryALTinterwordspacing}{\spaceskip=\fontdimen2\font plus
\BIBentryALTinterwordstretchfactor\fontdimen3\font minus
  \fontdimen4\font\relax}
\providecommand{\BIBforeignlanguage}[2]{{%
\expandafter\ifx\csname l@#1\endcsname\relax
\typeout{** WARNING: IEEEtran.bst: No hyphenation pattern has been}%
\typeout{** loaded for the language `#1'. Using the pattern for}%
\typeout{** the default language instead.}%
\else
\language=\csname l@#1\endcsname
\fi
#2}}
\providecommand{\BIBdecl}{\relax}
\BIBdecl

\bibitem{shen2023toward}
X.~Shen, J.~Gao, M.~Li, C.~Zhou, S.~Hu, M.~He, and W.~Zhuang, ``Toward
  immersive communications in {6G},'' \emph{Front. Comput. Sci.}, vol.~4, 2023.

\bibitem{peng2024semantic}
H.~Peng, Z.~Zhang, Y.~Liu, Z.~Su, T.~H. Luan, and N.~Cheng, ``Semantic
  communication in non-terrestrial networks: {A} future-ready paradigm,''
  \emph{IEEE Netw.}, vol.~38, no.~4, pp. 119--127, 2024.

\bibitem{tao2024wireless}
Z.~Tao, W.~Xu, Y.~Huang, X.~Wang, and X.~You, ``Wireless network digital twin
  for {6G}: Generative {AI} as a key enabler,'' \emph{IEEE Wirel. Commun.},
  vol.~31, no.~4, pp. 24--31, 2024.

\bibitem{yu2019cybertwin}
Q.~Yu, J.~Ren, Y.~Fu, Y.~Li, and W.~Zhang, ``Cybertwin: {An} origin of next
  generation network architecture,'' \emph{IEEE Wirel. Commun.}, vol.~26,
  no.~6, pp. 111--117, 2019.

\bibitem{sun2024knowledge}
R.~Sun, N.~Cheng, C.~Li, F.~Chen, and W.~Chen, ``Knowledge-driven deep learning
  paradigms for wireless network optimization in {6G},'' \emph{IEEE Netw.},
  vol.~38, no.~2, pp. 70--78, 2024.

\bibitem{mazumder2023dataperf}
M.~Mazumder, C.~Banbury, X.~Yao, B.~Karla{\c{s}}, W.~G. Rojas, S.~Diamos,
  G.~Diamos, L.~He, A.~Parrish, H.~R. Kirk \emph{et~al.}, ``{DataPerf}:
  {Benchmarks} for data-centric {AI} development,'' in \emph{NIPS}, 2023, New
  Orleans, LA, USA.

\bibitem{shen2021holistic}
X.~Shen, J.~Gao, W.~Wu, M.~Li, C.~Zhou, and W.~Zhuang, ``Holistic network
  virtualization and pervasive network intelligence for {6G},'' \emph{IEEE
  Commun. Surveys Tuts.}, vol.~24, no.~1, pp. 1--30, 2021.

\bibitem{zha2023data}
D.~Zha, Z.~P. Bhat, K.-H. Lai, F.~Yang, Z.~Jiang, S.~Zhong, and X.~Hu,
  ``Data-centric artificial intelligence: {A} survey,''
  \emph{arXiv:2303.10158}, 2023, [Online]. Available:
  https://arxiv.org/abs/2303.10158.

\bibitem{zhou2024digital}
C.~Zhou, J.~Gao, M.~Li, N.~Cheng, X.~Shen, and W.~Zhuang, ``Digital twin-based
  {3D} map management for edge-assisted device pose tracking in mobile {AR},''
  \emph{IEEE IoT J.}, vol.~11, no.~10, pp. 17\,812--17\,826, 2024.

\bibitem{ramakrishnan2002database}
R.~Ramakrishnan and J.~Gehrke, \emph{Database management systems}.\hskip 1em
  plus 0.5em minus 0.4em\relax McGraw-Hill, Inc., 2002.

\bibitem{ma2024one}
X.~Ma, L.~Luo, and Q.~Zeng, ``From one thousand pages of specification to
  unveiling hidden bugs: {Large} language model assisted fuzzing of {Matter}
  {IoT} devices,'' in \emph{USENIX Security}, 2024, Philadelphia, PA, USA.

\bibitem{wu2023characterizing}
F.~Wu, F.~Lyu, J.~Ren, P.~Yang, K.~Qian, S.~Gao, and Y.~Zhang, ``Characterizing
  internet card user portraits for efficient churn prediction model design,''
  \emph{IEEE Trans. Mobile Comput.}, vol.~23, no.~2, pp. 1735--1752, 2023.

\bibitem{larose2014discovering}
D.~T. Larose and C.~D. Larose, \emph{Discovering knowledge in data: {An}
  introduction to data mining}.\hskip 1em plus 0.5em minus 0.4em\relax John
  Wiley \& Sons, 2014, vol.~4.

\bibitem{ghorbani2020distributional}
A.~Ghorbani, M.~Kim, and J.~Zou, ``A distributional framework for data
  valuation,'' in \emph{ICML}, 2020, Virtual Conference.

\bibitem{han2020vivo}
B.~Han, Y.~Liu, and F.~Qian, ``Vivo: {Visibility}-aware mobile volumetric video
  streaming,'' in \emph{Proc. ACM MobiCom}, 2020, New York, NY, USA.

\end{thebibliography}

\bibliographystyle{IEEEtran}

\begin{IEEEbiographynophoto}{Conghao Zhou}
[S'19, M’22] received the Ph.D. degree in Electrical and Computer Engineering from the University of Waterloo, Canada. He is a Professor with the School of Telecommunications Engineering, Xidian University, China. His research interests include space-air-ground integrated networks, AI for networking, and immersive communications. Dr.~Zhou received the IEEE GLOBECOM'24 Best Paper Award and the IEEE PIMRC'23 Best Paper Award. 
\end{IEEEbiographynophoto}

\begin{IEEEbiographynophoto}{Shisheng Hu}
[S'19] received the B.Eng. and M.A.Sc. degrees from the University of Electronic Science and Technology of China (UESTC), Chengdu, China, in 2018 and 2021, respectively. He is currently pursuing the Ph.D. degree with the Department of Electrical and Computer Engineering, University of Waterloo, Waterloo, ON, Canada. His research interests include AI for wireless networks and networking for AI.
\end{IEEEbiographynophoto}

\begin{IEEEbiographynophoto}{Jie Gao}
[M'17, SM'21] received the Ph.D. degree from the University of Alberta. He worked as a Post-Doctoral Fellow at Toronto Metropolitan University, a Research Associate at the University of Waterloo, and an Assistant Professor at Marquette University. He is an Assistant Professor with the School of Information Technology, Carleton University, Ottawa, ON. His research interests include AI for networking and B5G/6G networks.
\end{IEEEbiographynophoto}

\begin{IEEEbiographynophoto}{Xinyu Huang}
[S'21] received the B.E. and M.S. degrees from Xidian University and Xi’an Jiaotong University, Xi’an, China, in 2018 and 2021, respectively. He is working toward the Ph.D. degree in Electrical and Computer Engineering at the University of Waterloo, Waterloo, ON, Canada. His research interests include digital agents, generative AI, and network resource management.
\end{IEEEbiographynophoto}

\begin{IEEEbiographynophoto}{Weihua Zhuang}
[M’93, SM’01, F’08] has been with the Department of Electrical and Computer Engineering, University of Waterloo, Canada, since 1993, where she is a University Professor and a University Research Chair in wireless communication networks. Her current research focuses on network architecture, algorithms and protocols, and service provisioning in future communication systems. She is an elected member of the Board of Governors (BoG) of the IEEE Vehicular Technology Society.
\end{IEEEbiographynophoto}

\begin{IEEEbiographynophoto}{Xuemin (Sherman) Shen}
[M’97, SM’02, F’09]~is a University Professor with the Department of E\&CE, University of Waterloo, Canada. His research focuses on network resource management, wireless network security, Internet of Things, 5G and beyond, and vehicular networks. Dr. Shen is the Past President of the IEEE ComSoc. He is a Canadian Academy of Engineering Fellow, a Royal Society of Canada Fellow, a Chinese Academy of Engineering Foreign Member, and an International Fellow of the Engineering Academy of Japan.
\end{IEEEbiographynophoto}

\end{document}